\documentstyle[preprint,pre,aps,epsfig,amssymb]{revtex}
\begin{document}
\draft

\title{Binary hard-sphere fluids near a hard wall}

\author{R. Roth\thanks{present address: University of Bristol, 
H.H. Wills Physics Laboratory, Royal Fort, Tyndall Avenue, Bristol BS8 1TL, UK}
and S. Dietrich}
\address{Fachbereich Physik, Bergische Universit\"at Wuppertal, D-42097
  Wuppertal, Germany}

\maketitle
\begin{abstract}
\noindent
By using the Rosenfeld density functional we determine the number density
profiles of both components of binary hard-sphere fluids close to a planar
hard wall as well as the corresponding excess coverage and surface tension. 
The comparison with published simulation data demonstrates that
the Rosenfeld functional, both its original version and sophistications 
thereof, is superior to previous approaches and
exhibits the same excellent accuracy as known from studies of the corresponding
one-component system.
\end{abstract}


\narrowtext

\section{Introduction}

A variety of experimental techniques has emerged which allow one to resolve the
inhomogeneous density distributions of fluids at interfaces, a subject which
enjoys broad scientific interest. In this context the ability to manufacture
highly monodisperse colloidal suspensions has turned out to be particular
useful as they provide the possibility to tune the effective interactions in
these systems such that, e.g., the colloidal particles closely resemble 
hard-sphere fluids \cite{Pusey91} . Since many of these experimental probes 
are indirect scattering techniques there is a substantial demand to guide 
them theoretically. Computer simulations and integral theories \cite{Hansen86}
are important tools of statistical physics to address these issues. Density 
functional theory (DFT) \cite{Evans79} has emerged as an additional approach 
which is capable to capture interfacial phase transitions and to sweep the 
thermodynamic and interaction parameter space of the system under 
consideration. The potential to combine these two possibilities poses already a
major challenge for the other techniques. If DFT acquires in addition the same
accuracy as the other two techniques, it could gain a clear competitive edge.

Although there is no recipe for constructing systematically a reliable DFT in
spatial dimensions $d\geq 2$, the constant flux of developments over many years
has led to a rather high level of sophistication. Among these theories for 
hard-sphere fluids, which act as paradigmatic systems and stepping-stones for 
more complicated models, the Rosenfeld functional has emerged as a particular 
powerful theory which resorts to the fundamental geometrical measures of the 
individual sphere \cite{Rosenfeld89}. For the standard test case of the highly
inhomogeneous density distribution of a one-component hard-sphere fluid near 
a planar hard wall, the predictions of the Rosenfeld functional are very close
to those of numerical simulations serving as benchmarks. For this case the 
mean square deviations [see, c.f., Eq.~(\ref{meanerror})] of Rosenfeld DFT 
results from the simulation data from Ref.~\cite{Groot87} are at most 
$1 \times 10^{-3}$ at high packing fractions, otherwise less than 
$3 \times 10^{-4}$.

Another virtue of the Rosenfeld functional is that is easily lends itself to 
the generalization to multi-component hard-sphere fluids. This opens the door 
to investigate rich new physical phenomena as particles of different size 
compete for interfacial positions \cite{Pagonabarraga00}. Even for the 
simplest multi-component system, the binary hard-sphere fluid, there are only 
relatively few theoretical studies which determine their structural properties
near a planar hard wall, using Monte Carlo simulations 
\cite{Tan89,Moradi89,Sokolowski90,Noworyta98}, integral equation theories 
\cite{Sokolowski90}, and various kinds
of density functional theory 
\cite{Moradi89,Sokolowski90,Noworyta98,Denton91,Leidl93,Patra99,Choudhury99,Kim99a},
as well as in spherical pores \cite{Kim99}.
Here we analyze this problem by using the corresponding Rosenfeld functional 
both in its original version \cite{Rosenfeld89} as well as for sophistications
thereof \cite{RSLT1,RSLT2}. By comparing these results with published 
simulation data \cite{Noworyta98} we assess to which extent the quantitative 
reliability of the Rosenfeld functional for the one-component hard-sphere 
fluid remains valid for the corresponding binary system. Moreover we determine
concentration profiles, the excess coverage, and the surface tension of the 
binary hard-sphere fluids at a hard wall. We describe the DFT in 
Sec.~\ref{DFT} and report our results in Sec.~\ref{results} followed by a 
summary and our conclusions in Sec.~\ref{conclusion}. The Appendix contains 
important technical details.

\section{Density functional theory} \label{DFT}

The Rosenfeld functional for the excess (over the ideal gas) Helmholtz free 
energy of a mixture of hard spheres with number density profiles
$\{\rho_i({\bf r})\},~i=1,\dots,N$, can be written as \cite{Rosenfeld89}
\begin{equation} \label{Fex}
\beta {\cal F}_{ex}[\{n_\alpha\}] = \int {\mathrm d}^3 r~ 
\Phi(\{ n_\alpha({\bf r}) \})
\end{equation}
which is a functional of the four scalar weighted densities 
$n_\alpha({\bf r})$ for the $N$-component mixture
\begin{equation} \label{weighted}
n_\alpha({\bf r}) = \sum_{i=1}^{N} \int {\mathrm d}^3 r'~ \rho_i({\bf r'})~ 
\omega_i^{(\alpha)} ({\bf r}-{\bf r}'),~\alpha=0,\dots,3,
\end{equation}
with $4 N$ scalar weight functions $\omega_i^{(\alpha)}$ and two 
three-component vector weighted densities ${\bf n}_\alpha({\bf r})$
\begin{equation}
{\bf n}_\alpha({\bf r}) =\sum_{i=1}^{N} \int {\mathrm d}^3 r'~ 
\rho_i({\bf r'})~ 
\mbox{\boldmath $\omega$}_i^{(\alpha)} ({\bf r}-{\bf r}'),~\alpha=1,2,
\end{equation}
with $2 N$ vector weight functions \mbox{\boldmath $\omega$}$_i^{(\alpha)}$. 
The weight functions contain {\em only} information about the fundamental 
geometrical measures of a single sphere of species $i$, namely its volume, 
surface area, and radius $R_i$, i.e., in particular they are independent of 
the density profiles. The explicit expressions for the weight functions are 
given in the Appendix. $\Phi(\{n_\alpha\})=\Phi_1+\Phi_2+\Phi_3$ is a function
of the weighted densities with \cite{Rosenfeld89}
\begin{eqnarray}
\Phi_1 & = & - n_0 \log(1-n_3), \\
\Phi_2 & = & \frac{n_1 n_2 - {\bf n}_1 \cdot {\bf n}_2}{1-n_3},
\end{eqnarray}
and
\begin{eqnarray}
\Phi_3 & = & \frac{\frac{1}{3} n_2^3 - n_2 {\bf n}_2 \cdot {\bf n}_2}
{8 \pi (1-n_3)^2} = \frac{n_2^3}{24 \pi (1-n_3)^2}~(1-3 \xi^2)\label{p3},
\end{eqnarray}
where $\xi({\bf r}) \equiv |{\bf n}_2({\bf r})|/n_2({\bf r})$, which is the 
ratio of the modulus of the vector weighted density ${\bf n}_2({\bf r})$ and 
the scalar weighted density $n_2({\bf r})$. We want to note that 
$\xi({\bf r}) \equiv 0$ in the bulk and is small for small inhomogeneities. 
While this original Rosenfeld functional describes very successfully the fluid
phase of a one-component hard-sphere system \cite{Rosenfeld99}, it fails 
to predict the freezing transition. This failure has been studied in detail in 
Refs.~\cite{RSLT1} and \cite{RSLT2}. For the freezing transition it turns out 
that the zero-dimensional limit of the functional, in which a small cavity can
accommodate only a single sphere, plays a key role. In a crystal the thermal 
vibrations around a lattice site can be interpreted as the motions in such a 
cavity formed by the neighboring spheres. Only if the statistical mechanics in
such a cavity is described properly by the density functional, the freezing 
transition is predicted correctly. This is not the case for the original 
Rosenfeld functional. This problem can be fixed by modifying slightly the 
contribution $\Phi_3$ [see  Eq.~(\ref{p3})] such that the freezing transition 
is predicted by the modified functional while at the same time, for lower 
packing fractions, the accuracy of the original functional in describing the 
inhomogeneous fluid is kept. The following modifications have been suggested 
\cite{RSLT1,RSLT2}:
\begin{equation} \label{asym}
\Phi_{3,q}=\frac{n_2^3}{24 \pi}~(1-\xi^2)^q
\end{equation}
with $q\geq 2$ and
\begin{equation} \label{int}
\Phi_{3,int} = \frac{n_2^3}{24 \pi}~(1 - 3\xi^2+2\xi^3).
\end{equation} 
The first suggestion, $\Phi_{3,q}$, is an antisymmetrized version of 
$\Phi_3$ in Eq.~(\ref{p3}) and the second, $\Phi_{3,int}$, interpolates
between $\Phi_3$ of Eq.~(\ref{p3}) and $\Phi_{3,0D}$ in the exact 
zero-dimensional limit
\begin{equation}
\Phi_{3,0D} = \frac{n_2^3}{24 \pi(1-n_3)^2}~\xi (1-\xi)^2.
\end{equation}
While the modified Rosenfeld function with $\Phi_{3,0D}$ does successfully 
predict the freezing transition of the one-component system, it leads to 
modified bulk properties and hence cannot describe the hard-sphere fluid 
as accurate as the original Rosenfeld functional. We note that the difference 
between $\Phi_3$ of the original Rosenfeld functional and both $\Phi_{3,q}$ 
with $q=3$ and $\Phi_{3,int}$ is of the order of ${\cal O}(\xi^3)$. Therefore
we expect the biggest differences between the various versions of the Rosenfeld
DFT to occur close to the wall where $\xi$ is largest.

Both the original Rosenfeld functional and the modifications corresponding to
Eqs.~(\ref{asym}) and (\ref{int}), i.e., the functionals that share common
bulk properties, are very successful and accurate for the 
one-component fluid. But far less is known for binary mixtures. While in this
latter respect in Ref.~\cite{Rosenfeld93} very good agreement between the 
density profiles obtained from the Rosenfeld functional and simulation data 
from Ref.~\cite{Tan89} has been mentioned, in a recent study \cite{Noworyta98} 
significant deviations between the Rosenfeld DFT results and simulations have 
been found. In this latter study an alternative but equivalent formulation 
\cite{Kierlik90,Phan93} of the original Rosenfeld functional has been applied.

Here we are interested in the equilibrium density profiles 
$\rho_{s,0}({\bf r})$ and $\rho_{b,0}({\bf r})$ of both the {\em s}mall and 
{\em b}ig components of binary hard-sphere mixtures close to a planar 
hard wall. To this end we freely minimize the functional
\begin{equation} \label{functional}
\Omega[\rho_s({\bf r}),\rho_b({\bf r})] = 
{\cal F}[\rho_s({\bf r}),\rho_b({\bf r})] 
+ \sum_{i=s,b}\int {\mathrm d}^3 r'~
\rho_i({\bf r}') \left( V_{ext}^i({\bf r}') - \mu_i \right)
\end{equation}
which is written in terms of the functional
\begin{equation}
{\cal F}[\rho_s({\bf r}),\rho_b({\bf r})] = {\cal F}_{id}[\rho_s({\bf r}),
\rho_b({\bf r})] + {\cal F}_{ex}[\rho_s({\bf r}),\rho_b({\bf r})]
\end{equation}
with the exactly known ideal gas contribution ${\cal F}_{id}$,
\begin{equation}
\beta {\cal F}_{id} = \sum_{i=s,b} \int {\mathrm d}^3 r'~ \rho_i({\bf r}')
\left( \ln(\lambda_i^3 \rho_i({\bf r}')) - 1 \right),
\end{equation}
with $\lambda_i$ the thermal wave length of species $i$. For the equilibrium
density profiles $\rho_{i,0}({\bf r})$, $i=s,b$, the functionals ${\cal F}$ and
$\Omega$ reduce to the Helmholtz free energy and the grand canonical
potential of the mixture, respectively; $\mu_s$ and $\mu_b$ are the chemical
potentials of the two species. The external potentials entering into
Eq.~(\ref{functional}) model the planar hard wall at $z=0$:
\begin{equation}
V_{ext}^{i}(z) = \left\{
\begin{array}{rl}
\infty, & z < R_i, \\
0, & \mbox{otherwise},
\end{array}
\right.
\end{equation}
$i=s,b$, with $z$ the normal distance from the wall. The external
potentials prevent the centers of spheres of species $i$ to approach the wall,
located at $z=0$, closer than $R_i$ in which case they are in contact.

In the absence of spontaneous symmetry breaking due to freezing, which we do 
not consider here, the profiles $\rho_{i,0}(z)$, $i=s,b$, depend only on the 
normal coordinate $z$ which simplifies the minimization of the functional.

Far away from the wall, i.e., in the bulk system, the vector weighted 
densities ${\bf n}_1$ and ${\bf n}_2$ and thus $\xi$ vanish. In this limit both
the original Rosenfeld functional and the two modifications corresponding
to Eqs.~(\ref{asym}) and (\ref{int}) reduce to the same bulk expression given 
by
\begin{equation} \label{bulk}
\Phi_{bulk}=-n_0 \ln(1-n_3) + \frac{n_1 n_2}{1-n_3} + \frac{n_2^3}
{24 \pi (1-n_3)^2}
\end{equation}
and hence they share the same bulk properties. We want to emphasize that as a 
consequence of this feature {\em all} versions of the Rosenfeld functional 
predict density profiles which show the {\em same} asymptotic decay towards 
the bulk value \cite{Evans94}. The weighted densities in the bulk limit are 
obtained by inserting the bulk densities $\rho_{i,bulk}:=\rho_{i,0}(z=\infty)$ 
into Eq.~(\ref{weighted}) yielding
\begin{eqnarray}
n_3 & = & \frac{4 \pi}{3} \sum_{i=s,b} R_i^3 \rho_{i,bulk} \equiv \sum_{i=s,b}
\eta_i, \label{bulkn3} \\
n_2 & = & 4 \pi \sum_{i=s,b} R_i^2 \rho_{i,bulk},\\
n_1 & = & \sum_{i=s,b} R_i \rho_{i,bulk},
\end{eqnarray}
and
\begin{eqnarray} \label{bulkn0}
n_0 & = & \sum_{i=s,b} \rho_{i,bulk}.
\end{eqnarray}
The equation of state following from Eq.~(\ref{bulk}),
\begin{equation} \label{pymix}
\beta p = \frac{n_0}{1-n_3} + \frac{n_1 n_2}{(1-n_3)^2}
+\frac{1}{12 \pi} \frac{n_2^3}{(1-n_3)^3},
\end{equation}
is the Percus-Yevick compressibility equation of state of the mixture 
\cite{Lebowitz64}. This expression is related to the contact values of the 
density profiles according to  the sum rule \cite{Rickayzen85}
\begin{equation} \label{sumrule}
\beta p = \sum_{i=s,b} \rho_i(z=R_i+0).
\end{equation}
This sum rule is respected by the Rosenfeld functional as by any 
weighted-density DFT \cite{Swol89} and therefore provides a test for the 
numerical accuracy of the calculations. In the following we suppress the 
subscript 0 which indicates equilibrium profiles as opposed to variational 
functions.

\section{Structural and thermodynamic properties} \label{results}

\subsection{Density profiles}

The number density profiles of both components of binary hard-sphere mixtures 
close to a planar hard wall are obtained by a free minimization of the 
functional given in Eq.~(\ref{functional}). We use the original Rosenfeld 
functional as well as the modified versions corresponding to Eq.~(\ref{int}) 
and to Eq.~(\ref{asym}) with $q=2$ and $q=3$. The systems considered here have
two different size ratios, $R_b:R_s=5:3$ and $R_b:R_s=3:1$, and various 
packing fractions $\eta_s$ and $\eta_b$ of the small and big spheres 
\cite{eta}, respectively. The resulting density profiles are compared with 
simulation data published in Ref.~\cite{Noworyta98}. In addition we calculate 
the local concentrations $\Phi_s(z)$ and $\Phi_b(z)$ of the small and big 
spheres, respectively, defined as
\begin{equation}
\Phi_i(z) = \frac{\rho_i(z)}{\rho_s(z)+\rho_b(z)},~i=s,b.
\end{equation}

We find excellent agreement between the density profiles of both components 
obtained by density functional theory and the simulation data for all systems 
under consideration. This holds for all versions of the Rosenfeld functional 
analyzed here. While at low total packing fractions $\eta=\eta_s+\eta_b$
the density functional theory results for all versions of the Rosenfeld 
functional are practically equivalent, small deviations among the results from 
different versions of the functional appear at larger values of $\eta$, i.e., 
for $\eta \gtrsim 0.3$. We quantify the degree of agreement between our DFT 
density profiles $\rho_i(z)$ and the simulation data from 
Ref.~\cite{Noworyta98}, available as data points 
$(z_j, \rho_i^{sim}(z_j))$, $j=1\dots N_i^{sim}$ and $i=s,b$, by determining 
the mean square deviations $\bar E_i$, $i=s,b$, defined as
\begin{equation} \label{meanerror}
\bar E_i = \frac{1}{N_i^{sim}} \sum_{j=1}^{N_i^{sim}} \left( 
\frac{\rho_i^{sim}(z_j)-\rho_i(z_j)} {\rho_{i,bulk}} \right)^2.
\end{equation}
We find that $\bar E_s$ and $\bar E_b$ are at most $5 \times 10^{-4}$ and
$6 \times 10^{-3}$, respectively, for all versions of the Rosenfeld 
functional. However, since the statistical errors in the simulation data are 
comparable with or even larger than the differences between the density 
profiles obtained by different version of the Rosenfeld DFT this approach does 
not enable us to determine which of the various versions is the most accurate 
one. 

To illustrate the agreement between the DFT and the simulation data, in 
Fig.~\ref{fig:prof1} we show the density profiles of the small spheres (a) and 
of the big spheres (b) for $\eta_s=0.0607$ and $\eta_b=0.3105$ and a size 
ratio $R_b:R_s=5:3$. The symbols ($\square$) denote the simulation data from 
Ref.~\cite{Noworyta98} and the solid line is obtained by the original 
Rosenfeld functional. The dotted lines denote coarse grained densities 
$\bar \rho_i^{(j)}$, $i=s,b$ and $j=0,1$, defined as
\begin{equation} \label{coarse}
\bar \rho_i^{(j)} = \frac{1}{z_i^{(j+1)}-z_i^{(j)}} \int 
\limits_{z_i^{(j)}}^{z_i^{(j+1)}} {\mathrm d} z~ 
\frac{\rho_i(z)}{\rho_{i,bulk}},
\end{equation}
with $z_i^{(0)}\equiv0$ and $z_i^{(j>0)}$ the position of the $j$th minimum 
\cite{details35}. All details of the density profiles found in the simulations
are reproduced very accurately by the density functional theory. The 
oscillatory behavior, i.e., the amplitudes, the phases, and the decay of the 
oscillations as obtained by DFT agree excellently with the simulations. The 
total packing fraction of the system $\eta=\eta_s+\eta_b=0.3712$ is already 
rather high, giving rise to the pronounced structure of the density profiles. 
The corresponding concentration profiles $\Phi_s(z)$ and $\Phi_b(z)$ of the 
small and big spheres, respectively, are shown in Fig.~\ref{fig:conc1}. These 
concentration profiles demonstrate that, apart from the purely geometrical 
constraints, near the wall the big particles are enriched and the small 
particles depleted. This is in line with the expectation based on the 
attractive depletion potential near a hard wall of a single big sphere 
immersed in a fluid of small spheres \cite{Goetzelmann98}. Correlation effects
reverse this relative distribution in the second layer and restore it in the 
third.

As mentioned above, small differences between the DFT results corresponding to
the various versions of the Rosenfeld functional can be found for these values
of $\eta$. In order to be able to resolve these small differences magnified 
parts of the density profiles from Fig.~\ref{fig:prof1} are shown in 
Fig.~\ref{fig:diff1} together with the simulation data from 
Ref.~\cite{Noworyta98}($\square$). The solid lines in Fig.~\ref{fig:diff1}
correspond to the original Rosenfeld functional, the dotted lines correspond
to the interpolated version [Eq.~(\ref{int})] whereas the dashed and 
dashed-dotted lines correspond to the antisymmetrized version 
[Eq.~(\ref{asym})] with $q=2$ and $q=3$, respectively. All DFT results are 
very close to the simulation data. However, the deviations between the 
simulations and the antisymmetrized functional with $q=2$ seem to be 
systematically the largest.

In Figs.~\ref{fig:prof2} and \ref{fig:diff2} we show the density profiles of a
binary mixture with size ratio $R_b:R_s=3:1$. The packing fraction of the 
small spheres is $\eta_s=0.0047$ and that of the big spheres is 
$\eta_b=0.3859$ so that the total packing fraction $\eta=\eta_s+\eta_b=0.3906$
is again rather high. Therefore there is a strong spatial variation of the 
density profiles. The agreement between DFT (solid line) and simulations 
($\square$) is again found to be excellent for both the density profile of the
small spheres (a) and of the big spheres (b). The dotted lines denote the 
coarse grained densities as defined in Eq.~({\ref{coarse})\cite{details31}. The
concentrations profiles of the small and big spheres, corresponding to these 
density profiles are shown in Fig.~\ref{fig:conc2}. For this larger size ratio
the anti-correlated behavior of $\Phi_s(z)$ and $\Phi_b(z)$ is even more 
pronounced than for the smaller ratio discussed in Fig.~\ref{fig:conc1} and 
strongly locked in without additional fine structure such as the double peak 
appearing in Fig.~\ref{fig:conc1}.

In addition we test the numerical accuracy of our calculations by means of the
sum rule given in Eq.~(\ref{sumrule}). In Table~\ref{tab:sum35} the sum of the
contact values of the binary mixture with size ratio $R_b:R_s=5:3$ for various
packing fractions is compared with two equations of state. $\beta p_{PY}^c$ 
denotes the Percus-Yevick compressibility equation of state 
[Eq.~(\ref{pymix})], to which the Rosenfeld functional reduces in the bulk 
limit, and $\beta p_{MCSL}$ corresponds to the more accurate 
Mansoori-Carnahan-Starling-Leland equation of state \cite{Mansoori71}, which 
represents a generalization to a mixture of the very accurate 
Carnahan-Starling equation of state \cite{Carnahan69} for the one-component 
fluid. The very good agreement between the contact values and $\beta p_{PY}^c$
demonstrates the high accuracy of our numerical procedure. However, at higher 
packing fractions, $\beta p_{PY}^c$ deviates from the more accurate equation 
of state $\beta p_{MCSL}$. The same analysis of our results for a binary 
mixture with size ratio $R_b:R_s=3:1$ is summarized in Table~\ref{tab:sum31}. 

Equation~(\ref{sumrule}) represents a sum rule which must be fulfilled by the
density profiles as obtained by any of the density functionals considered here.
However, no corresponding rules are available for the individual contact 
values. We find that for all versions of the Rosenfeld functional under 
consideration here, the sum rule is respected equally well. However, the 
individual contact values may differ. This statement is in line with the 
expectation that the biggest differences between the various versions of the 
Rosenfeld functional occur in a region where $\xi$ is large, i.e., close to 
the wall, and is substantiated in Table~\ref{tab:contact35} for the binary 
mixture with size ratio $R_b:R_s=5:3$ and in Table~\ref{tab:contact31} for
the size ratio $R_b:R_s=3:1$.

Vested with this confidence in our numerical procedures we are now able to 
comment on the comparison of our DFT results for the original Rosenfeld 
functional with those obtained by earlier DFT studies 
\cite{Moradi89,Sokolowski90,Noworyta98,Denton91,Leidl93,Patra99,Choudhury99,Kim99a}.
We find that DFT approaches other than the Rosenfeld DFT predict the structure 
of the density profiles of a binary hard-sphere mixture near a planar 
hard wall only qualitatively 
\cite{Moradi89,Sokolowski90,Denton91,Leidl93,Patra99,Choudhury99,Kim99a}, 
whereas the Rosenfeld functional yields quantitatively reliable predictions. 
This is demonstrated by Figs.~\ref{fig:prof1} and 
\ref{fig:diff1}--\ref{fig:diff2} and the very small values of the mean square 
deviations $\bar E_s$ and $\bar E_b$ [Eq.~(\ref{meanerror})]. The predictions 
of the Rosenfeld functional agree excellently in all details with the 
simulation results from Ref.~\cite{Noworyta98}. The deviations between the DFT
results of Ref.~\cite{Noworyta98}, calculated with the Rosenfeld functional, 
and their own simulations originate most likely from numerically problems of 
the iteration procedure used in Ref.~\cite{Noworyta98}. This suspicion is 
further supported by the fact that the DFT density profiles shown in 
Ref.~\cite{Noworyta98} seem to violate the sum rule Eq.~(\ref{sumrule}). Thus 
we are led to the conclusion that the deviations between the Rosenfeld DFT 
results and the simulation data reported in Ref.~\cite{Noworyta98} most likely
are artifacts generated by numerical problems in implementing the iteration 
procedure used in Ref.~\cite{Noworyta98} for solving the Euler-Lagrange 
equations corresponding to the Rosenfeld density functional. Therefore the 
doubts raised in Ref.~\cite{Noworyta98} about the performance of the Rosenfeld
DFT for binary hard-sphere mixtures are not justified. We conclude that the 
Rosenfeld DFT exhibits the same high accuracy in predicting density 
distributions of binary hard-sphere mixtures as for the one-component 
hard-sphere fluid.

\subsection{Excess adsorption and surface tension}

One of the virtues of DFT is that based on the knowledge of the local 
structural properties $\rho_i(z)$, $i=s,b$, it is straightforward to calculate
also thermodynamic properties such as the excess adsorptions $\Gamma_i$ and 
the surface tension $\gamma$. Here we determine these quantities near a 
hard wall for a binary hard-sphere fluid whose components exhibits a size 
ratio $R_b:R_s=3:1$. Our analysis is confined to the fluid phase of the 
mixture; the phase boundary for freezing is estimated from the bulk phase 
diagrams presented in Refs.~\cite{Dijkstra98}. 

The excess adsorption of species $i$, $i=s,b$, is defined as
\begin{eqnarray} \label{coverage}
\Gamma_i & = & \int \limits_0^\infty {\mathrm d}z~ (\rho_i(z) - 
\rho_{i,bulk}) \\
& = & \int \limits_{\sigma_i/2}^\infty {\mathrm d}z~ (\rho_i(z) - 
\rho_{i,bulk}) -
\frac{\sigma_i}{2} \rho_{i,bulk} \nonumber.
\end{eqnarray}
This definition of the excess adsorption differs from the definition used 
in Ref.~\cite{Goetzelmann96} as well as from that used in 
Ref.~\cite{Henderson84}. To recover the results for the excess adsorption of
a one-component hard-sphere fluid in Ref.~\cite{Goetzelmann96} and 
Ref.~\cite{Henderson84} one has to subtract from and add to our results, 
respectively, the constant $\frac{\sigma_i}{2} \rho_{i,bulk}$.
These differences originate from different choices for the position of the 
wall. While for the one-component fluid there is no preference for any of 
these definitions, our choice used here appears to be particularly suited for 
a mixture because independent of the diameter $\sigma_i$ of species $i$ the 
integrals in Eq.~(\ref{coverage}) start at the same lower bound for all 
species, namely at the position of the physical wall. We use the same 
definition for the position of the wall to determine the surface tension. 
Because the excess adsorptions follow from integrating over oscillatory 
functions, they depend very sensitively on the precise structure of the 
density profiles and require very accurate calculations. Moreover, near the 
phase boundary for freezing the original Rosenfeld functional yields values 
for $\Gamma_s$ and $\Gamma_b$ which differ significantly from those obtained 
from the modifications of the Rosenfeld functional. Thus, unless stated 
otherwise, we have determined the excess adsorption by using the modified 
Rosenfeld functional corresponding to Eq.~(\ref{int}), which is known to
capture the freezing transition of the one-component hard-sphere fluid.

In Fig.~\ref{fig:covers} we show the excess adsorptions $\Gamma_s$ of the 
small spheres as function of the packing fractions $\eta_s$ and $\eta_b$; note
that $\Gamma_s(\eta_s = 0,\eta_b)\equiv 0$. For a fixed packing fraction of the
small spheres $\eta_s$, the excess adsorption of the small spheres increases
upon increasing $\eta_b$. The reason for this is, that the increasing packing
effects of the big spheres associated with large values of $\eta_b$ enforces 
also the packing of the small spheres, giving rise to a strongly enhanced 
contact value and very pronounced structures in the density profile of the 
small spheres. For very small $\eta_s$, i.e., $\eta_s \leq 0.1$, and large 
$\eta_b$ we find that $\Gamma_s$ can become positive. As function of $\eta_b$,
$\Gamma_s$ exhibits a turning point for any fixed value of $\eta_s$.

The excess adsorption $\Gamma_b$ of the big spheres is shown in 
Fig.~\ref{fig:coverb}. For constant $\eta_s$ and increasing $\eta_b$, 
$\Gamma_b$ decreases. But due to the same mechanism as described above, 
$\Gamma_b$ increases upon increasing $\eta_s$ for constant packing fractions 
of the big spheres $\eta_b$. The square symbols ($\square$) in 
Fig.~\ref{fig:coverb} denote simulation data for the excess adsorption of a 
one-component hard-sphere fluid near a hard wall at packing fractions 
$\eta_s$=0.3680, 0.4103, and 0.4364, respectively, taken from 
Ref.~\cite{Henderson84}. Whereas the simulation data for the two smaller
packing fractions agree very well with our DFT results, that for the highest 
value of $\eta_s$ clearly deviates from the DFT prediction. Thus from this
comparison it remains unclear to which extent the simulation data and the
DFT results are reliable close to freezing.

In order to illustrate the large differences between the excess adsorptions 
calculated from different functionals, in Fig.~\ref{fig:gsb_comp} we show the 
excess adsorption $\Gamma_s$ and $\Gamma_b$ calculated from the original 
Rosenfeld functional together with those calculated by using the modified 
functional corresponding to Eq.~(\ref{int}). The packing fraction of the small 
spheres is $\eta_s=0.15$. Only for small packing fractions of the big spheres, 
i.e., far away from freezing, we find good agreement between the results 
obtained from the different functionals. Close to the phase boundaries 
\cite{boundaries} the 
differences become large, which is indeed surprising at first glance. However, 
the main contribution of the excess adsorption stems from the vicinity of the 
wall where $\xi$ is large and the different versions of the Rosenfeld 
functional are expected to differ. These differences were already indicated in
the previous subsection by the behavior of the individual contact values 
$\sigma_i^3 \rho_i(R_i+0)$, $i=s,b$, of the density profiles of the small and 
big spheres, respectively. For large distances from the wall the density 
profiles exhibit decaying oscillations and therefore are not expected to 
contribute essentially to the excess adsorption. Moreover, {\em all} versions 
of the Rosenfeld functional display a common characteristic decay because they 
share the same bulk properties \cite{Evans94} so that the contributions to the 
excess adsorption far away from the wall are very similar for the original 
Rosenfeld functional and its modifications.  

The grand potential $\Omega$ of a system in contact with a wall
\begin{equation}
\Omega = \Omega_{bulk} + \Omega_{surf}
\end{equation}
decomposes into a bulk contribution, $\Omega_{bulk} = - p V$, given by the 
bulk pressure $p$ in the system times the volume $V$ occupied by fluid 
particles, and a surface contribution, $\Omega_{surf} = \gamma A$, which is 
the surface tension $\gamma$ times the surface area $A$ of the wall. Scaled 
particle theory (SPT) provides an approximate expression for $\gamma$ for a 
one-component hard-sphere fluid \cite{Reiss60} as well as a generalization to 
hard-sphere mixtures \cite{Lebowitz64} close to a planar hard wall. The 
surface tension of a one-component hard-sphere fluid within SPT is well tested
and turns out to provide reliable results as compared with both DFT 
calculations \cite{Goetzelmann96} and simulations \cite{Heni99}. In 
Ref.~\cite{Henderson87} a fit to simulation results of the surface tension of 
a one-component hard-sphere fluid at a planar hard wall is presented, which 
gives quasi-exact results and closely resembles the SPT expression.

In terms of the weighted bulk densities $n_0,\dots,n_3$, defined in
Eqs.~(\ref{bulkn3})-(\ref{bulkn0}), the SPT approximation for the surface 
tension of a hard-sphere mixture close to a planar hard wall \cite{Lebowitz64} 
can be written as
\begin{equation} \label{spt}
\beta \gamma_{SPT} = \frac{n_1}{1-n_3} + \frac{1}{8 \pi} 
\frac{n_2^2}{(1-n_3)^2}.
\end{equation}
This expression reduces to the one-component SPT approximation of the surface
tension when it is evaluated for the one-component bulk weighted densities. In 
this latter case the surface tension can be expressed solely in terms of the 
packing fraction $\eta$ of this single component.

Within the Rosenfeld functional the surface tension $\gamma$ of a binary 
hard-sphere mixture at a planar hard wall follows from the equilibrium
density profiles $\rho_i(z)$, $i=s,b$, as obtained in the previous subsection: 
\begin{eqnarray}
\beta \gamma & = & \frac{\beta \Omega + \beta p V}{A} \nonumber \\
& = & \lim_{L\to\infty} \left[ \beta p L +
\int \limits_0^L {\mathrm d} z~ \left\{\Phi[\rho_s(z),\rho_b(z)] +
\sum_{i=s,b} \rho_i(z) ( V_{ext}^i(z) - \mu_i) \right\} \right].
\end{eqnarray}

In contrast to the strong dependence of the results of the excess adsorption
on the choice of functional, we find that the original Rosenfeld functional
as well as its modifications predict very similar results for the surface 
tension in the whole range of packing fractions studied here. Our results for 
the surface tension of a binary hard-sphere mixture with size ratio 
$R_b:R_s=3:1$, calculated within the original Rosenfeld functional, are shown 
in Fig.~\ref{fig:gamma}. The deviation between these results and those 
obtained by the modifications of the Rosenfeld functional are at most 3\% and 
deviate from the predictions of the SPT at most by 10\%.

\section{Summary and conclusions} \label{conclusion}

Based on the Rosenfeld density functional we have analyzed structural and
thermodynamic properties of binary hard-sphere mixtures near a hard wall with
the following main results:

\begin{enumerate}
\item Figures~\ref{fig:prof1} and \ref{fig:prof2} demonstrate that the 
structure of both density profiles $\rho_i(z)$, $i=s,b$, of a binary 
hard-sphere mixture close to a planar hard wall as obtained by the original 
Rosenfeld functional is in excellent agreement with simulation results for 
size ratios $R_b:R_s=5:3$ and $R_b:R_s=3:1$, respectively. The high level of 
agreement between our DFT results and the simulation data from 
Ref.~\cite{Noworyta98} is confirmed quantitatively by small mean square 
deviations $\bar E_s$ and $\bar E_b$ defined in Eq.~(\ref{meanerror}). In 
terms of these quantities all versions of the Rosenfeld functional are 
practically of the same quality. Only at high packing fractions rather small 
differences between the results obtained by different versions of the 
Rosenfeld DFT become visible like those shown in Figs.~\ref{fig:diff1} and 
\ref{fig:diff2}.
\item The concentration profiles (see Fig.~\ref{fig:conc1} and 
Fig.~\ref{fig:conc2}) calculated from the density profiles confirm the 
depletion picture: the small spheres are depleted from regions close to the 
wall while the big spheres are enriched.
\item The numerical accuracy of our calculations is demonstrated in 
Tables~\ref{tab:sum35} and \ref{tab:sum31} by the high degree at which the sum 
rule Eq.~(\ref{sumrule}), which relates the sum of the contact values of both 
density profiles with the equation of state, is respected. The sum rule, 
however, makes no prediction for the individual contact values and we find in 
Tables~\ref{tab:contact35} and \ref{tab:contact31} that each version of the 
Rosenfeld functional takes a different route to satisfy the sum rule.
\item Using the modified Rosenfeld functional corresponding to Eq.(\ref{int})
we have calculated the excess adsorption of the small spheres 
$\Gamma_s(\eta_s,\eta_b)$  (see Fig.~\ref{fig:covers}) and of the big spheres
$\Gamma_b(\eta_s,\eta_b)$  (see Fig.~\ref{fig:coverb}) as functions of the 
packing fractions $\eta_s$ and $\eta_b$ for a binary hard-sphere mixture with 
size ratio $R_b:R_s=3:1$. We find that these quantities depend very 
sensitively on the accuracy of the numerical calculations and, as can be seen 
in Fig.~\ref{fig:gsb_comp}, they differ significantly from the excess 
adsorption calculated by the original Rosenfeld functional.
\item The surface tension of a binary hard-sphere mixture with size ratio 
$R_b:R_s=3:1$ close to a planar hard wall is shown in Fig.~\ref{fig:gamma}. We
find that all versions of the Rosenfeld functional give results which are in 
good agreement with the prediction of scaled particle theory [Eq.~(\ref{spt})].
\end{enumerate}

From these results we conclude that the class of Rosenfeld functionals yields 
quantitatively reliable descriptions of interfacial structures in binary 
hard-sphere fluids. We expect that the same level of reliability also holds 
for multi-component hard-sphere fluids. 

The excess adsorptions $\Gamma_s$ and $\Gamma_b$ of the small and big spheres
emphasize the differences between the various versions of the Rosenfeld 
functional most. In order to decide whether the original Rosenfeld functional 
or whether its modifications predict these quantities more accurately, 
additional simulation data of the excess adsorption in a binary hard-sphere 
fluid are needed.

\acknowledgments

It is a pleasure to thank Bob Evans for many stimulating discussions. We want 
to thank Dr. Noworyta for providing us with his simulation data.

\appendix

\section{Calculation of the weighted densities}

Within the minimization procedure of the Rosenfeld functional the weighted 
densities $n_\alpha$ and ${\bf n}_\alpha$ have to be calculated repeatedly. 
Therefore it is necessary to optimize these calculations with respect to both 
computational speed and numerical accuracy. 

The weight functions of the Rosenfeld functional are given by
\begin{eqnarray}
\omega_i^{(3)}({\bf r}) & = & \Theta(|{\bf r}|-R_i) \label{w3}, \\
\omega_i^{(2)}({\bf r}) & = & \delta(|{\bf r}|-R_i) \label{w2},
\end{eqnarray}
and
\begin{eqnarray}
\mbox{\boldmath $\omega$}_i^{(2)}({\bf r}) & = & \frac{{\bf r}}{|{\bf r}|}~
\delta(|{\bf r}|-R_i)
\label{wv2}
\end{eqnarray}
with the Heaviside function $\Theta$ and the Dirac delta function $\delta$.
The remaining scalar weight functions are proportional to $\omega_i^{(2)}$:
$\omega_i^{(1)}=\omega_i^{(2)}/(4 \pi R_i)$ and 
$\omega_i^{(0)}=\omega_i^{(2)}/(4 \pi R_i^2)$. The first vector weight 
function is collinear with {\boldmath $\omega$}$_i^{(2)}$:
{\boldmath $\omega$}$_i^{(1)}$={\boldmath $\omega$}$_i^{(2)}/(4 \pi R_i)$.

In order to calculate the weighted densities integrals $I_i^{(\alpha)}$ of the
type
\begin{equation}
I_i^{(\alpha)} = \int {\mathrm d}^3 r' \rho_i({\bf r}) 
\omega_i^\alpha({\bf r}-{\bf r}')
\end{equation}
have to be evaluated. For these convolution type integrals one can exploit the
symmetry properties of the density profiles. For the present geometry the 
weighted densities can be written as
\begin{equation} \label{weightz}
n_\alpha(z) = \sum_{i=s,b} \int_{-R_i}^{R_i} {\mathrm d} z' \rho_i(z+z')
\bar\omega_i^{(\alpha)}(z')
\end{equation}
with $s$ and $b$ for small and big, respectively, and with {\em reduced} 
weight functions $\bar\omega_i^{(\alpha)}$ which are functions of
$z$ only:
\begin{eqnarray}
\bar\omega_i^{(3)}(z) & = & \pi (R_i^2-z^2),\\
\bar\omega_i^{(2)}(z) & = & 2 \pi R_i,
\end{eqnarray}
and
\begin{eqnarray}
\mbox{\boldmath $\bar \omega$}_i^{(2)}(z) & = & 2 \pi z {\bf e}_z
\end{eqnarray}
with the unit vector ${\bf e}_z$ in $z$-direction. The relations between these
and the remaining weight functions are the same as for the original weight 
functions. The integrals in Eq.~(\ref{weightz}) are one-dimensional 
convolutions which can be calculated faster and more accurate in Fourier space 
than in real space. By introducing the Fourier transforms of the density 
profiles,
\begin{equation}
\hat\rho_i(k) = {\cal FT}(\rho_i(z)),
\end{equation}
and those of the weight functions,
\begin{equation}
\hat\omega_i^{(\alpha)}(k) = {\cal FT}(\bar\omega_i^{(\alpha)}(z)),
\end{equation}
the weighted densities can be expressed as
\begin{equation}
n_\alpha(z) = {\cal FT}^{-1}\left(\sum_{i=s,b} \hat\rho_i(k)
\hat\omega_i^\alpha(k)\right).
\end{equation}
This route of calculation offers the important advantage that the numerical 
calculation of these convolutions can be speed up significantly by applying 
Fast-Fourier-Transform (FFT) methods. Moreover it turns out that calculations 
of convolutions in real space depend more sensitively on the grid size 
$\Delta z$ to be used for discretization than those in Fourier space. We 
expect that the reason for this is that the FFT algorithm interpolates between 
data points with trigonometrical functions. To overcome this problem in real 
space a sophisticated integration scheme would have be applied or a very small
grid size would have to be chosen. Both remedies additionally slow down the 
numerical calculation in real space.

The results presented in this appendix are applicable if the density profiles 
depend on the $z$ coordinate only. However, similar results can be obtained if
the density profiles have radial symmetry.

\begin{figure}
\centering\epsfig{file=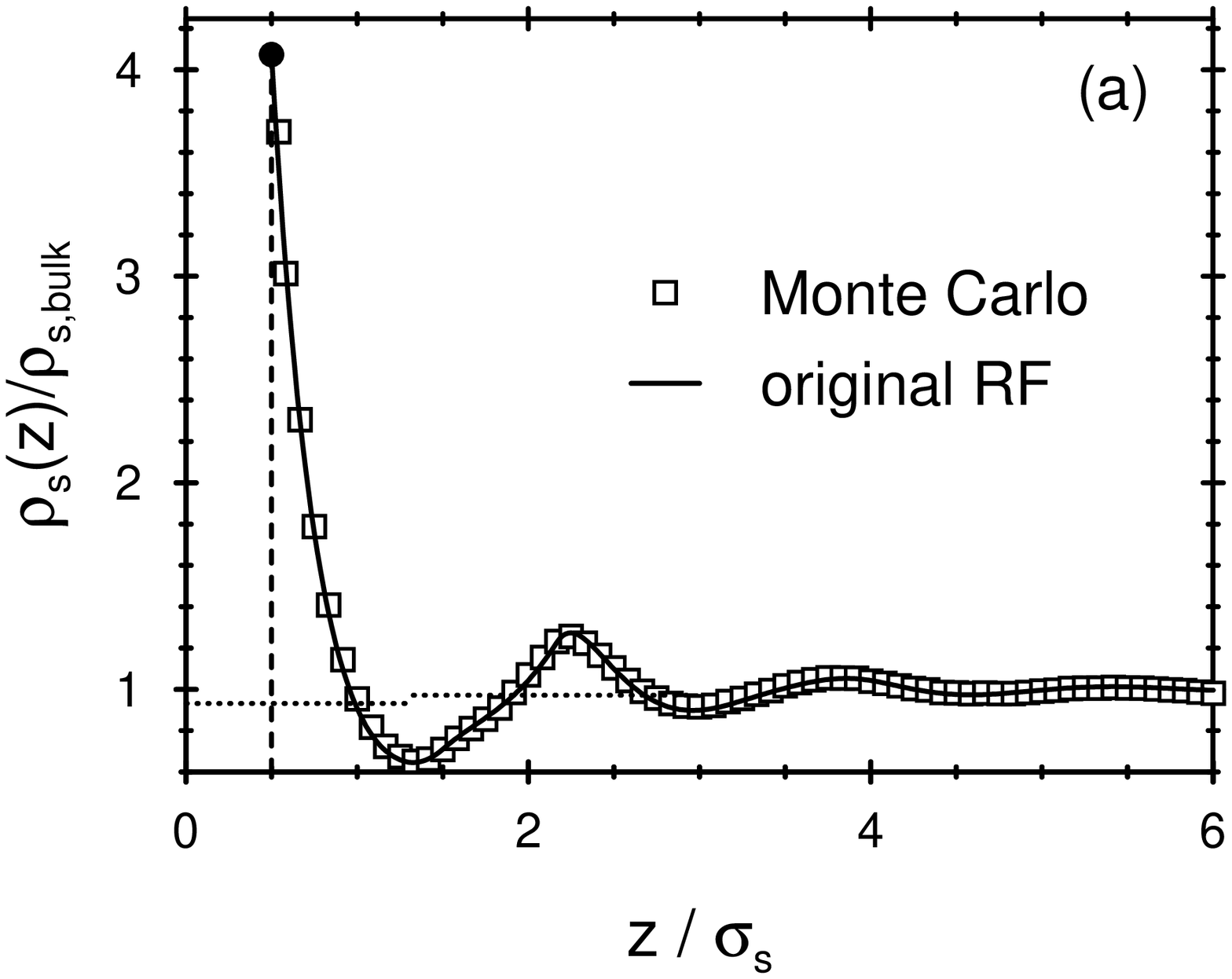,bbllx=10,bblly=60,bburx=550,bbury=460,
width=11cm}
\vspace{0.5cm}

\centering\epsfig{file=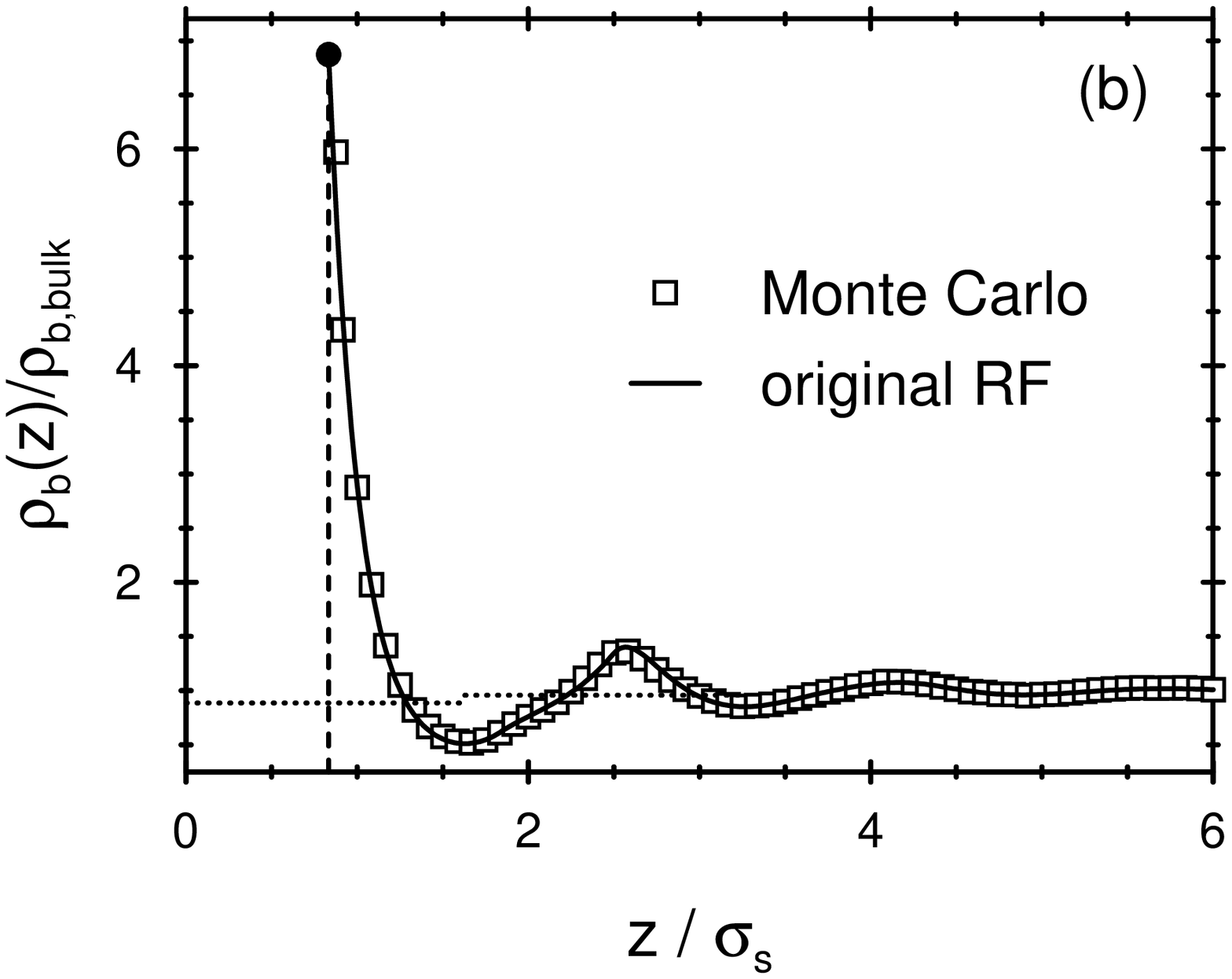,bbllx=10,bblly=60,bburx=550,bbury=460,
width=11cm}
\vspace{0.5cm}
\caption{The density profiles of the small spheres (a) and the big spheres (b)
of a binary hard-sphere mixture with size ratio $R_b:R_s=5:3$ and packing 
fractions $\eta_s=0.0607$ and $\eta_b=0.3105$ close to a planar hard wall
(dashed line). $\sigma_s=2 R_s$ is the diameter of the small spheres. The 
solid lines correspond to results obtained by a free minimization of the 
original Rosenfeld functional and the symbols ($\square$) denote simulation 
data from Ref.~\protect\cite{Noworyta98}. The full dots indicate the values at
contact. The positions of the extrema of $\rho_s(z)$ and $\rho_b(z)$ are very 
close to each other if they are measured from the corresponding position of 
contact, i.e., $z=R_s=\frac{1}{2} \sigma_s$ and $z=R_b=\frac{5}{6} \sigma_s$, 
respectively. The dotted lines correspond to the coarse grained density 
profiles defined in Eq.~(\ref{coarse}). They indicate that in spite of the 
high contact values the wall actually leads to a slight net depletion for both 
species in the first layer.}
\label{fig:prof1}
\end{figure}
\newpage

\begin{figure}
\centering\epsfig{file=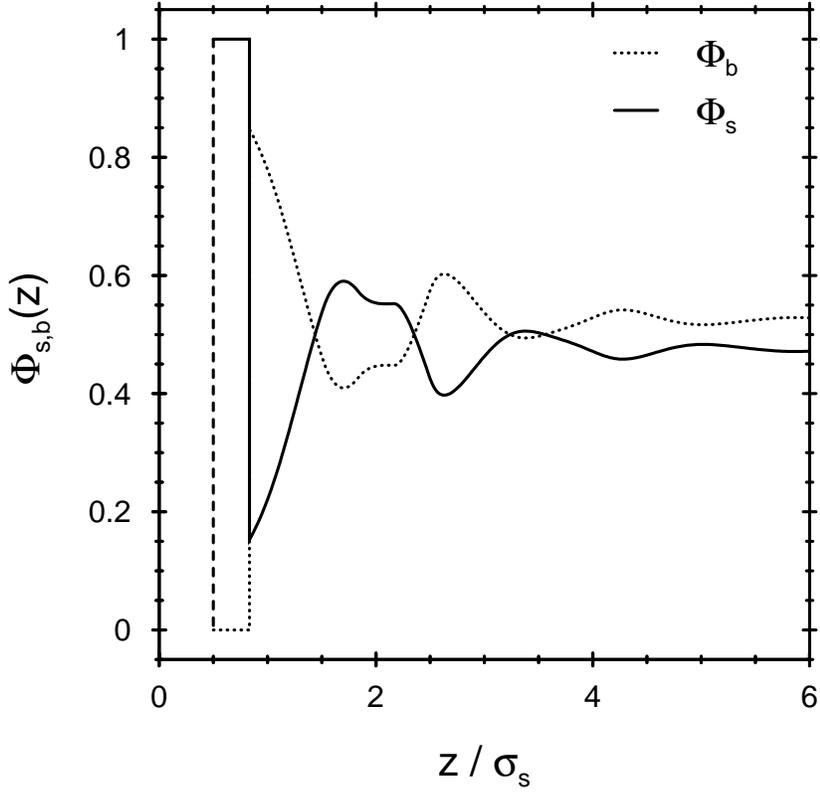,bbllx=10,bblly=60,bburx=550,bbury=560,
width=11cm}
\vspace{0.5cm}
\caption{The concentration profiles $\Phi_s(z)$ and $\Phi_b(z)$ of the small 
spheres and big spheres, respectively, corresponding to the density profiles in
Fig.~\ref{fig:prof1}. For geometric reasons $\rho_i(z \leq R_i) = 0$ with
$R_s=\frac{1}{2} \sigma_s$ and $R_b=\frac{5}{6} \sigma_b$. Therefore $\Phi_s$ 
and $\Phi_b$ are defined only for $z \geq \sigma_s$ and $\Phi_b(R_s<z<R_b)=0$ 
and $\Phi_b(R_s<z<R_b)=1$.}
\label{fig:conc1}
\end{figure}
\newpage

\begin{figure}
\centering\epsfig{file=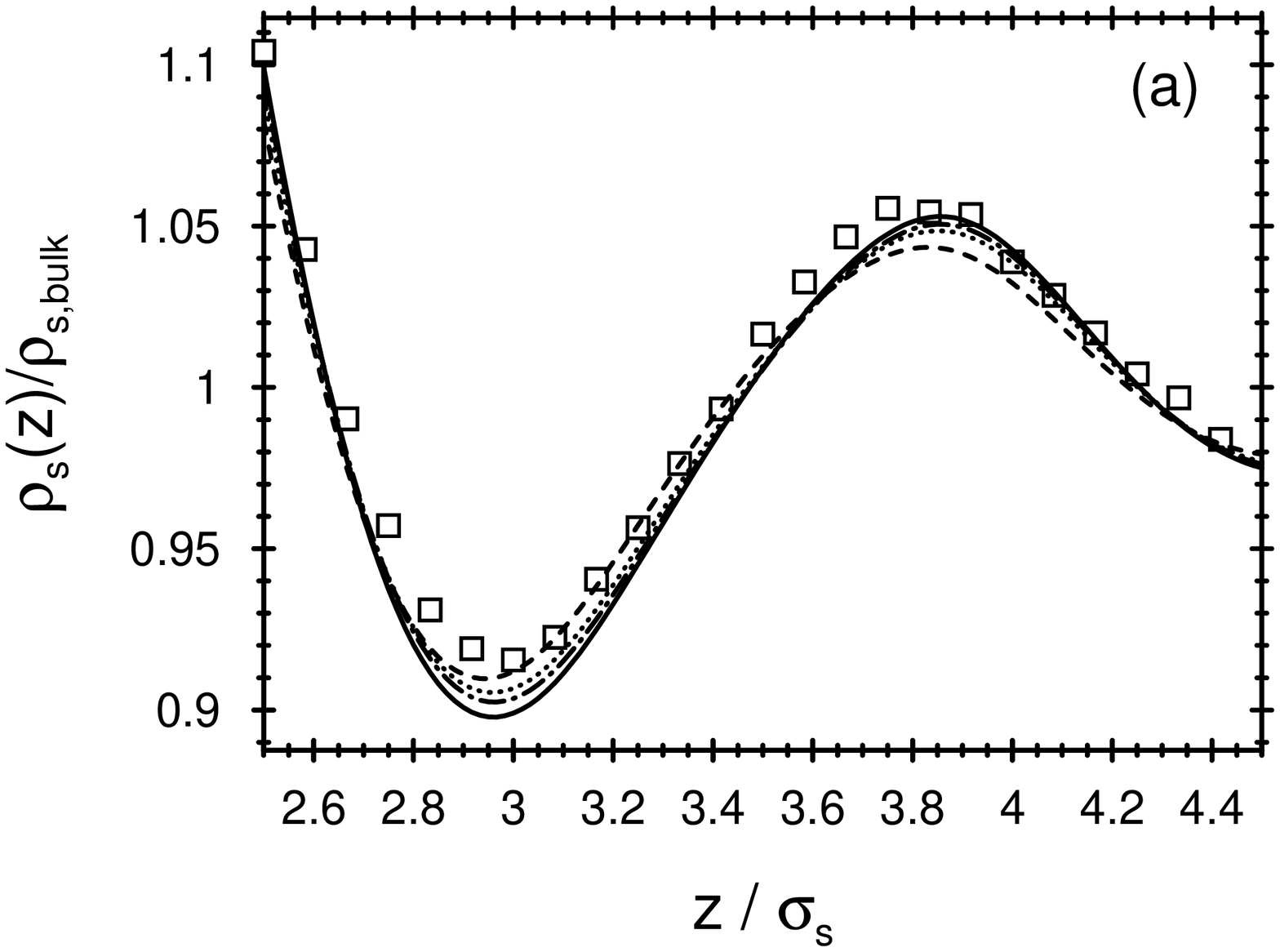,bbllx=10,bblly=60,bburx=550,bbury=460,
width=11cm}
\vspace{0.5cm}

\centering\epsfig{file=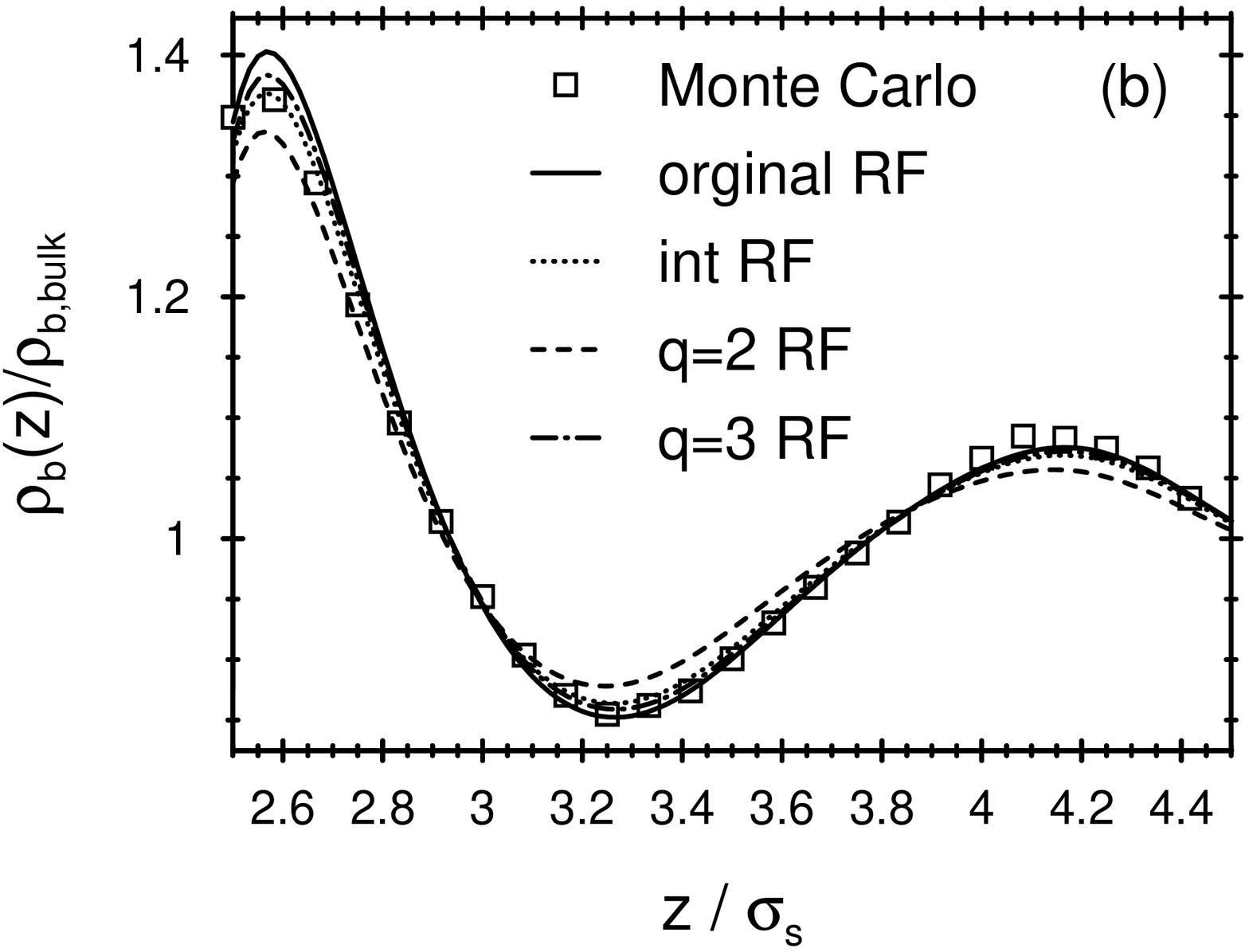,bbllx=10,bblly=60,bburx=550,bbury=460,
width=11cm}
\vspace{0.5cm}
\caption{In order to resolve the small differences in the density profiles of 
the small (a) and big (b) spheres between the DFT results obtained by the 
various versions of the Rosenfeld functional, parts of the density profiles 
shown in Fig.~\ref{fig:prof1} are magnified here. The solid lines correspond 
to the original functional, whereas the dotted, dashed, and dashed-dotted 
lines correspond the interpolated version (int RF), and the antisymmetrized 
modification with $q=2$ and $q=3$, respectively. The parameters
are the same as in Fig.~\ref{fig:prof1}.}
\label{fig:diff1}
\end{figure}

\begin{figure}
\centering\epsfig{file=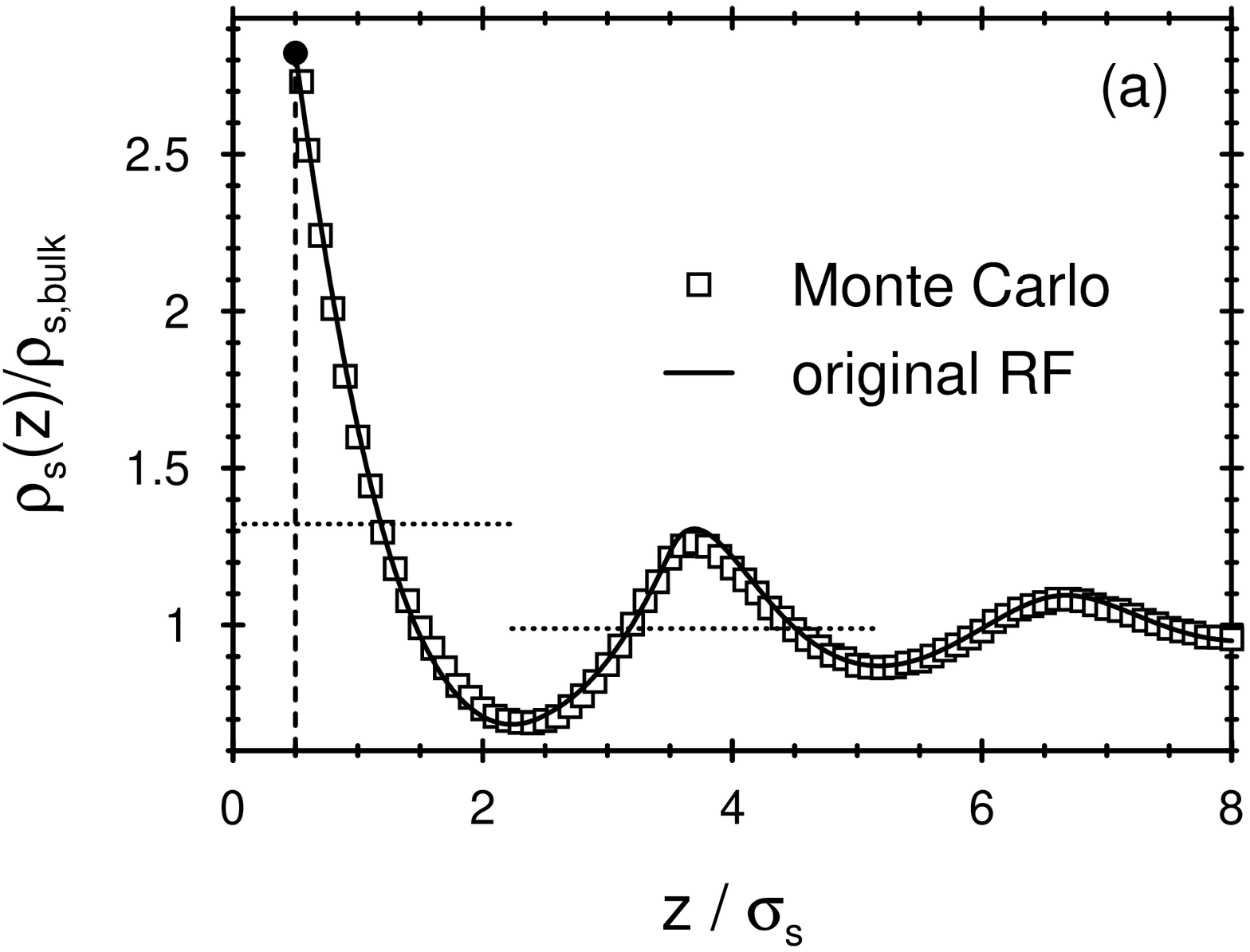,bbllx=10,bblly=60,bburx=550,bbury=460,
width=11cm}
\vspace{0.5cm}

\centering\epsfig{file=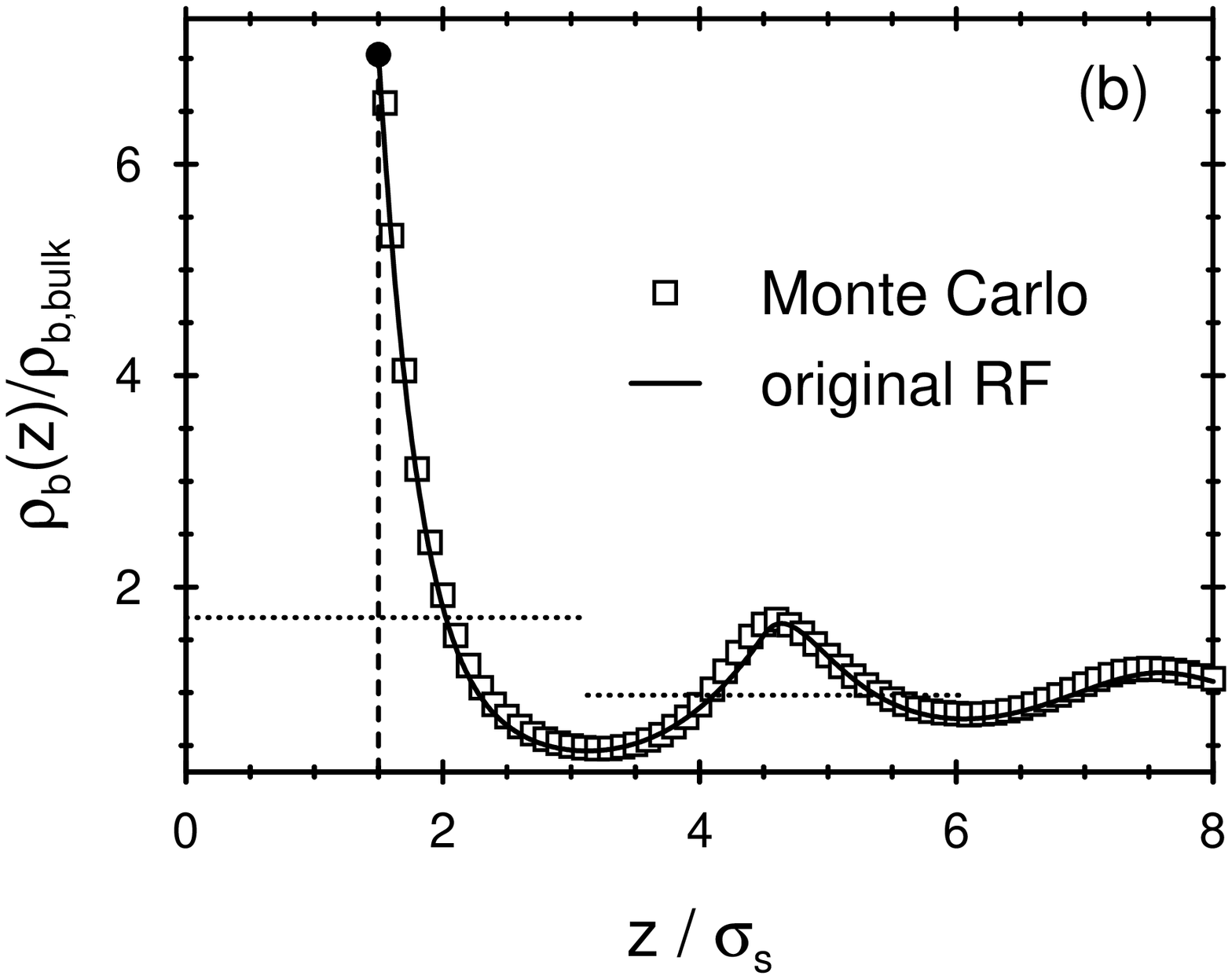,bbllx=10,bblly=60,bburx=550,bbury=460,
width=11cm}
\vspace{0.5cm}
\caption{The density profiles of the small spheres (a) and of the big spheres 
(b) of a binary hard-sphere mixture with size ratio $R_b:R_s=3:1$ and packing 
fractions $\eta_s=0.0047$ and $\eta_b=0.3859$ close to a planar hard wall. The 
solid lines correspond to results obtained by a free minimization of the 
original Rosenfeld functional. The simulation data ($\square$) are taken from
Ref.~\protect\cite{Noworyta98}. As in Fig.~\ref{fig:prof1} the positions of the
extrema of $\rho_s(z)$ and $\rho_b(z)$ are very close to each other if they are
measured from the corresponding position of contact, i.e., 
$z=R_s=\frac{1}{2}\sigma_s$ and $z=R_b=\frac{3}{2} \sigma_s$, respectively. 
The dotted lines correspond to the coarse grained density profiles defined in 
Eq.~(\ref{coarse}). Different from Fig.~\ref{fig:prof1}, here the net density 
of both species is clearly enhanced near the wall.}
\label{fig:prof2}
\end{figure}
\newpage

\begin{figure}
\centering\epsfig{file=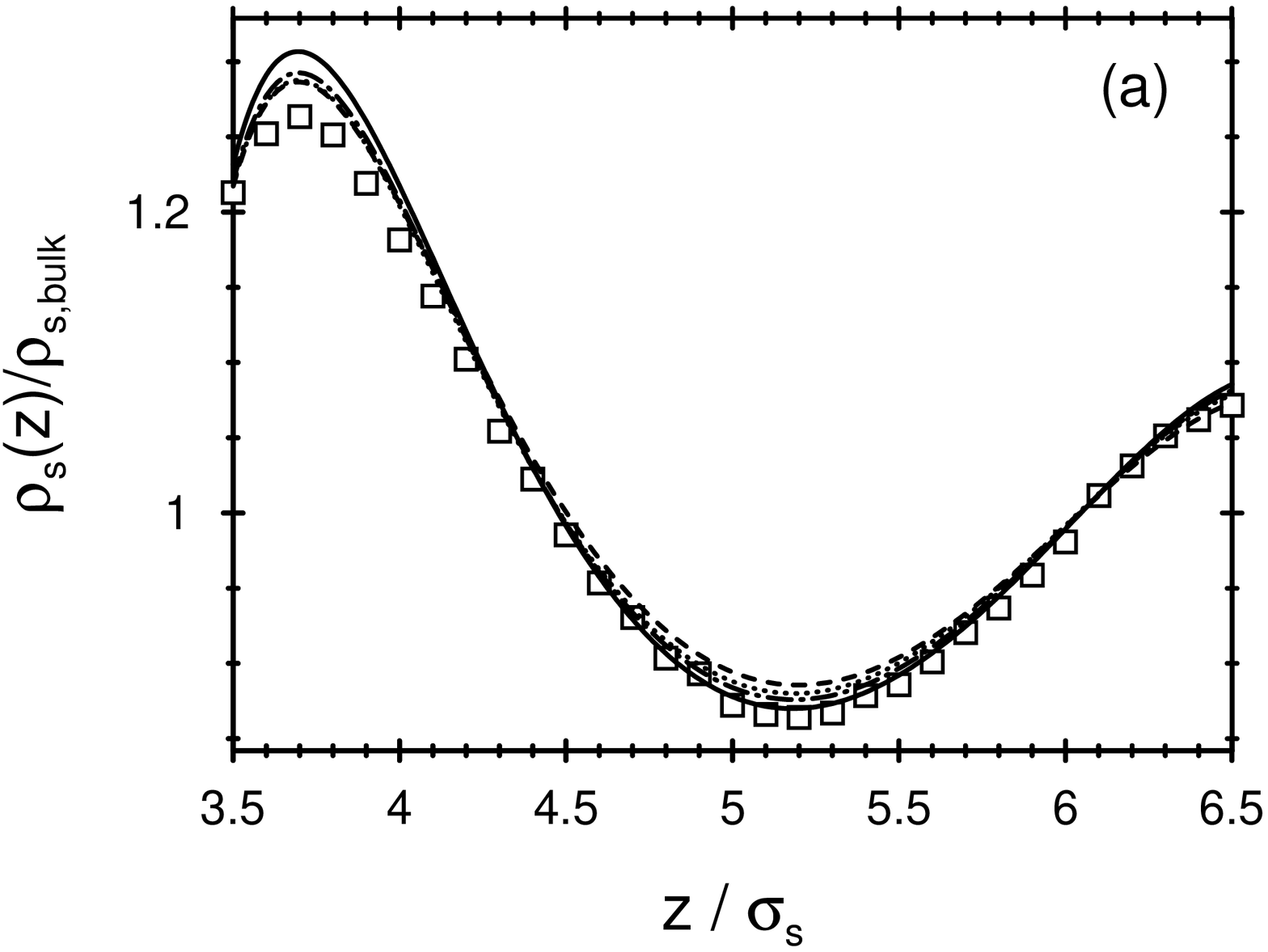,bbllx=10,bblly=60,bburx=550,bbury=460,
width=11cm}
\vspace{0.5cm}

\centering\epsfig{file=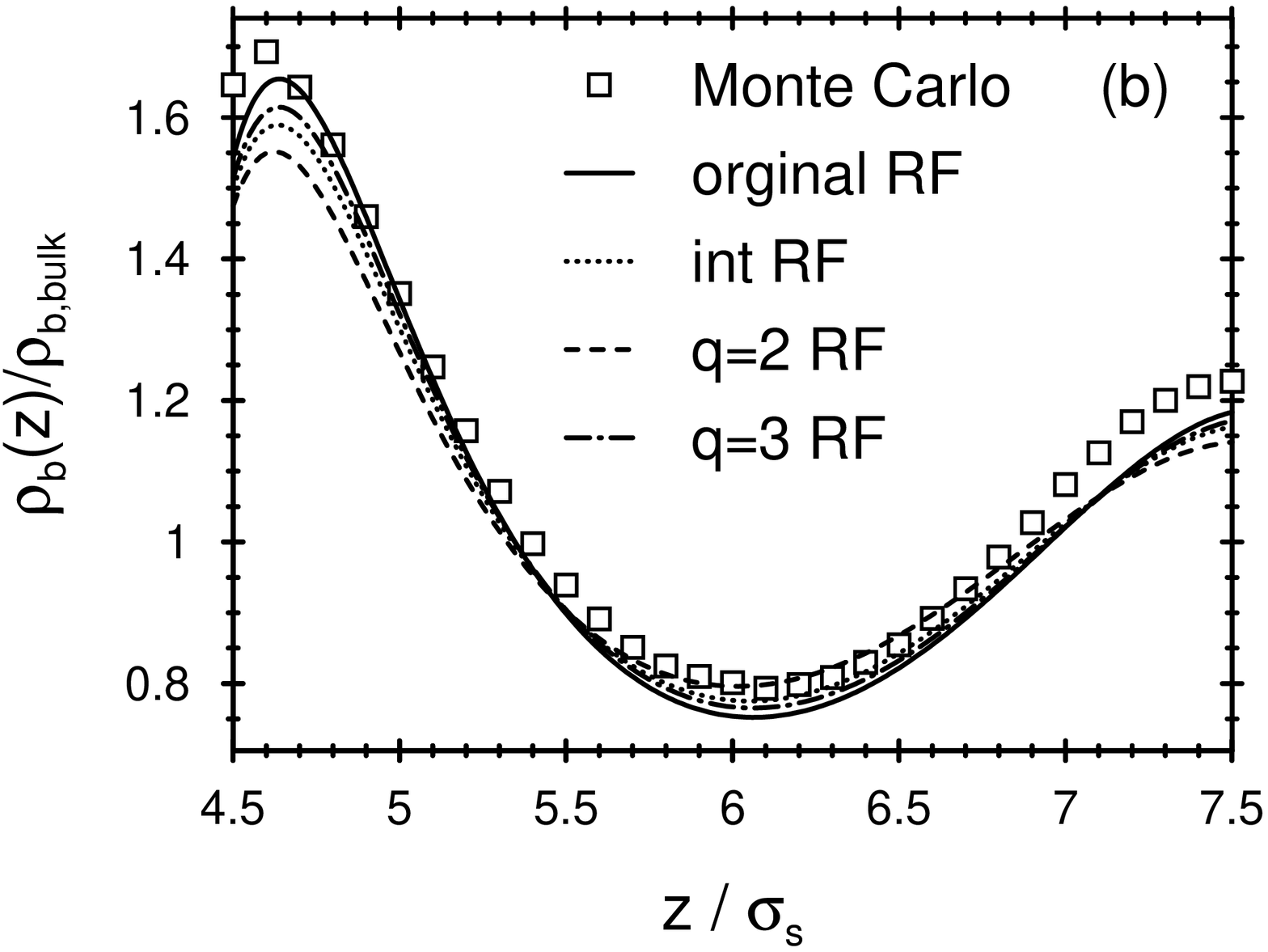,bbllx=10,bblly=60,bburx=550,bbury=460,
width=11cm}
\vspace{0.5cm}
\caption{In order to highlight the small differences of the DFT results 
obtained by different versions of the Rosenfeld functional, here small parts 
of the density profiles shown in Fig.~\ref{fig:prof2} are magnified.}
\label{fig:diff2}
\end{figure}

\begin{figure}
\centering\epsfig{file=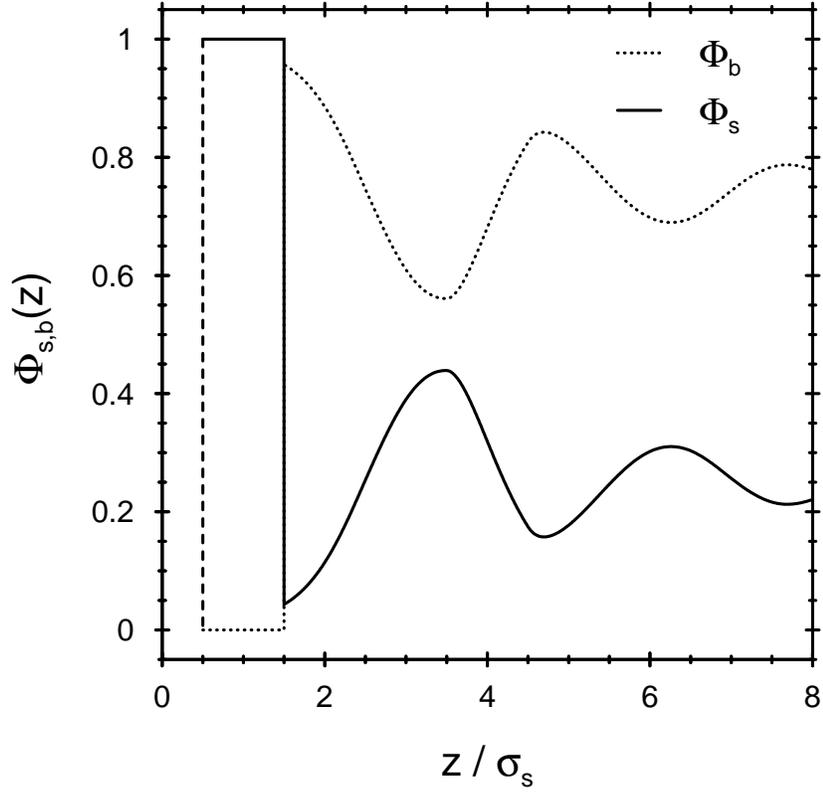,bbllx=10,bblly=60,bburx=550,bbury=560,
width=11cm}
\vspace{0.5cm}
\caption{The concentration profiles $\Phi_s(z)$ and $\Phi_b(z)$ of the small 
spheres and big spheres, respectively, corresponding to the density profiles of
Fig.~\ref{fig:prof2}. $R_s=\frac{1}{2} \sigma_s$ and $R_b=\frac{3}{2} \sigma_s$
so that due to geometric constraints $\Phi_b(R_s<z<R_b)=0$ and 
$\Phi_s(R_s<z<R_b)=1$. The anti-correlated behavior of the two profiles is 
strongly locked in.}
\label{fig:conc2}
\end{figure}
\newpage

\begin{figure}
\centering\epsfig{file=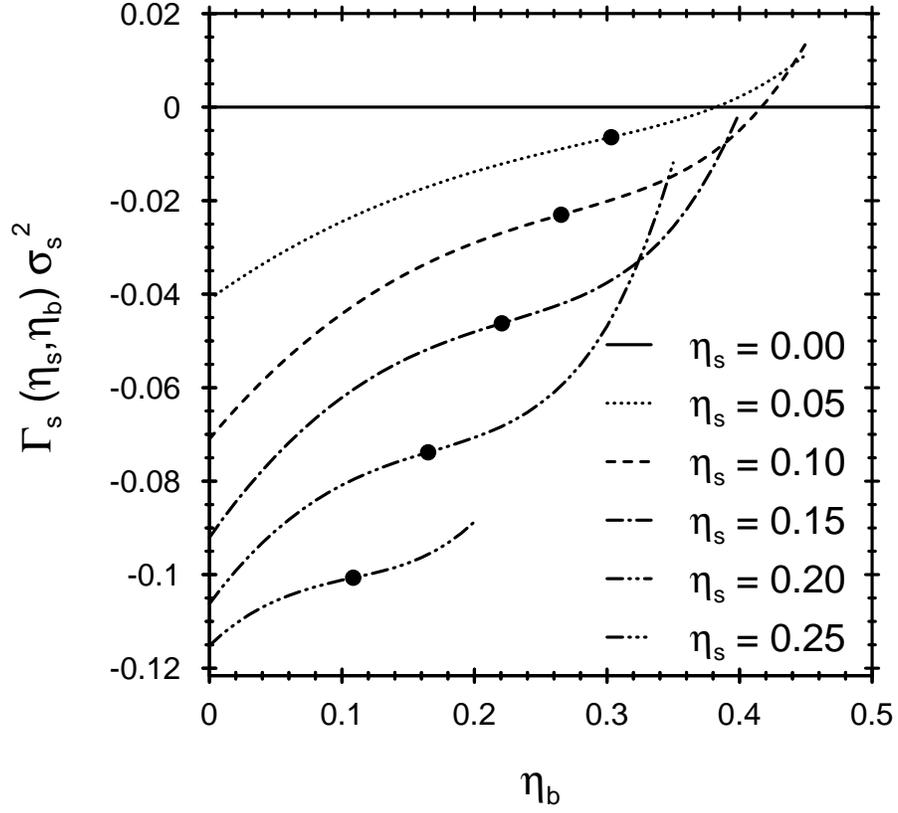,bbllx=0,bblly=60,bburx=530,bbury=580,
width=11cm}
\vspace{0.5cm}
\caption{The excess adsorption $\Gamma_s$ of the small spheres near a planar 
hard wall as function of the packing fractions $\eta_s$ and $\eta_b$ in the
fluid phase \protect\cite{boundaries}. The dots ($\bullet$) denote turning 
points.}
\label{fig:covers}
\end{figure}
\newpage

\begin{figure}
\centering\epsfig{file=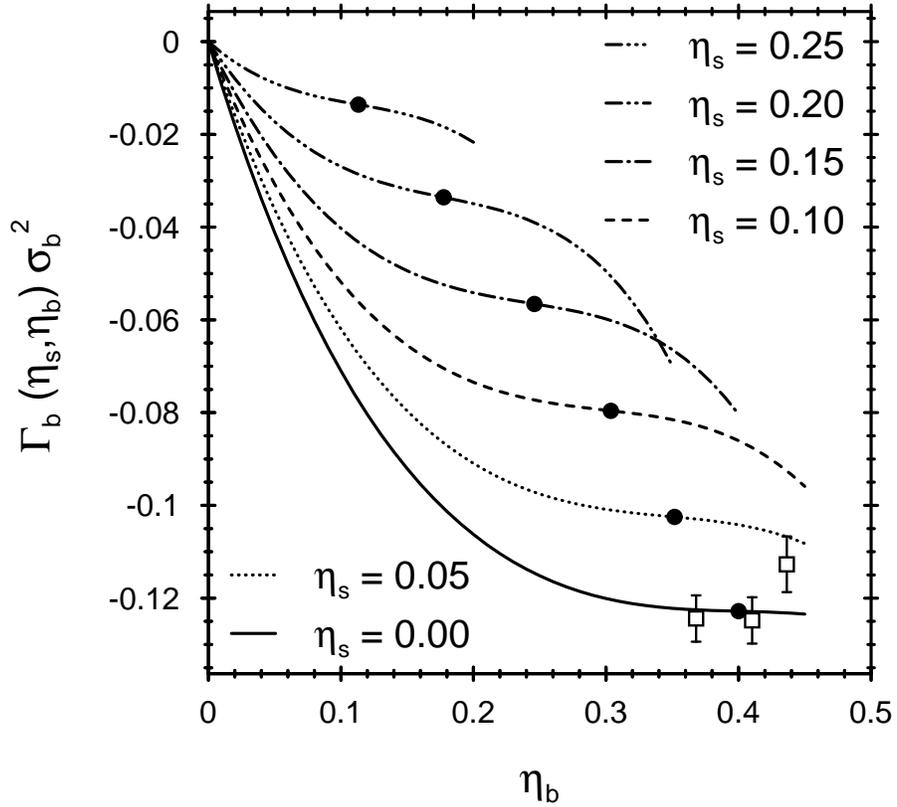,bbllx=0,bblly=60,bburx=530,bbury=580,
width=11cm}
\vspace{0.5cm}
\caption{The excess adsorption $\Gamma_b$ of the big spheres near a 
planar hard wall as function of the packing fractions $\eta_s$ and $\eta_b$
in the fluid phase \protect\cite{boundaries}. The dots ($\bullet$) denote 
turning points. The square symbols ($\square$) denote simulation results for 
the excess adsorption of a one-component hard-sphere fluid taken from 
Ref.~\protect\cite{Henderson84}.}
\label{fig:coverb}
\end{figure}

\begin{figure}
\centering\epsfig{file=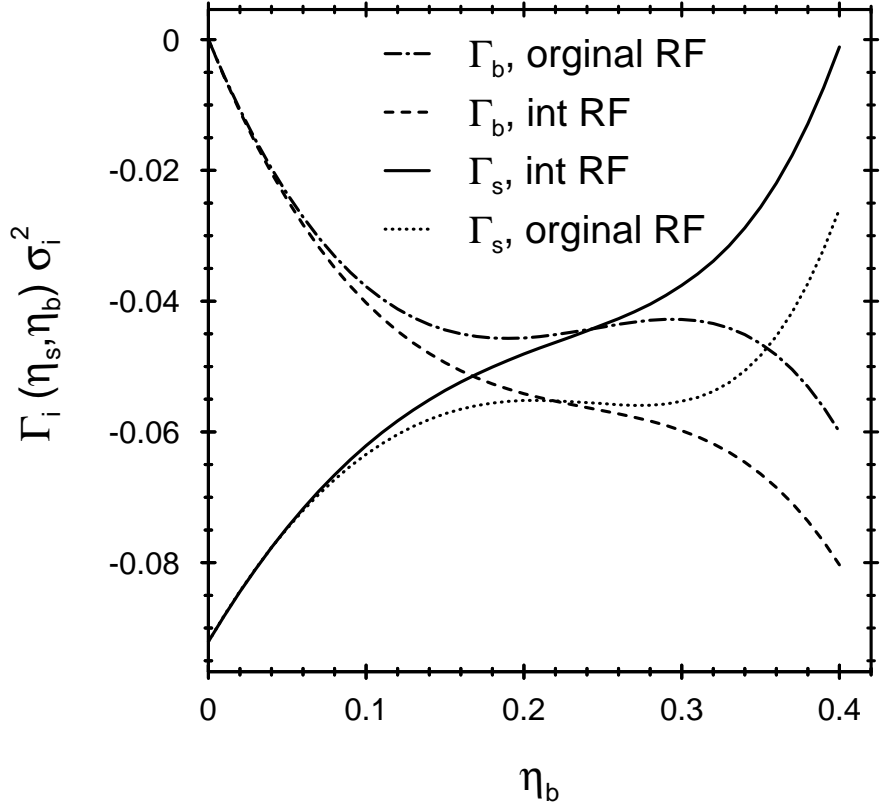,bbllx=0,bblly=60,bburx=530,bbury=580,
width=11cm}
\vspace{0.5cm}
\caption{Excess adsorptions $\Gamma_s$ and $\Gamma_b$ calculated with the 
original Rosenfeld functional compared with those obtained from its 
modification corresponding to Eq.~(\ref{int}). The packing fraction of the 
small spheres is $\eta_s=0.15$.}
\label{fig:gsb_comp}
\end{figure}
\newpage

\begin{figure}
\centering\epsfig{file=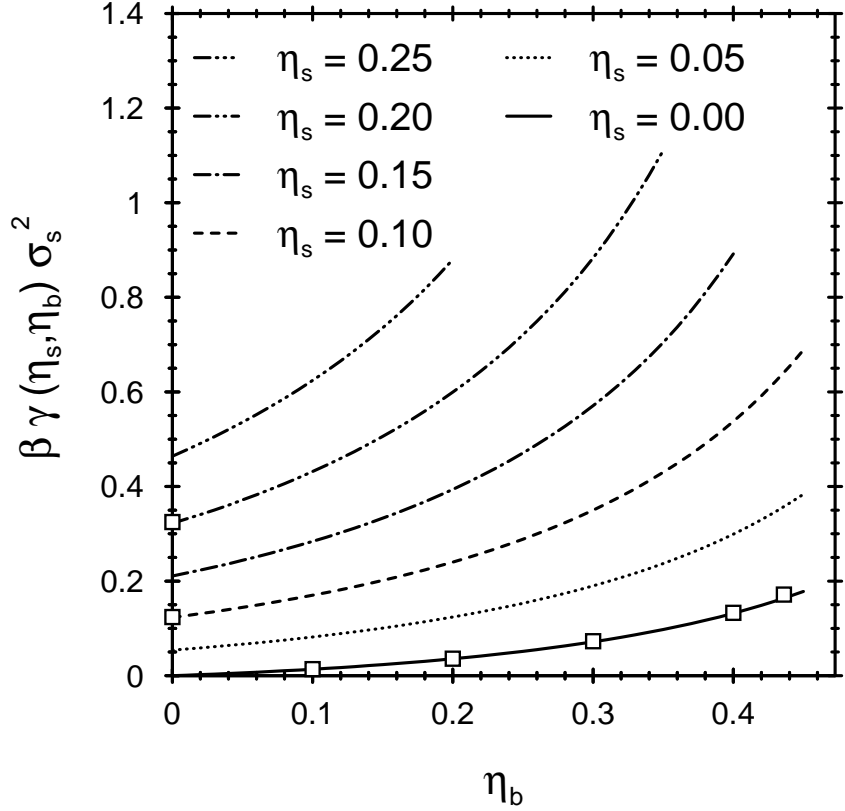,bbllx=0,bblly=60,bburx=530,bbury=580,
width=11cm}
\vspace{0.5cm}
\caption{The surface tension $\beta \gamma(\eta_s,\eta_b)$ of a binary 
hard-sphere mixture in the fluid phase \protect\cite{boundaries} with size 
ratio $R_b:R_s=3:1$ as function of the packing fractions $\eta_s$ and 
$\eta_b$, calculated from the original Rosenfeld functional (lines). For 
$\eta_s$=0 the hard-sphere fluid freezes at $\eta_b$=0.494. The square symbols
($\square$) denote simulation results from Ref.~\protect\cite{Heni99} for the 
one-component hard-sphere fluid.}
\label{fig:gamma}
\end{figure}

\begin{table}
\caption{Contact values at a planar hard wall of a binary hard-sphere mixture 
with size ratio $R_b:R_s=5:3$ for various packing fractions $\eta_s$ and 
$\eta_b$. The comparison of the sum of the contact values with the 
Percus-Yevick compressibility equation of state, which underlies the Rosenfeld
functional, tests the accuracy of our numerical procedures. For small total 
packing fractions $\eta=\eta_s+\eta_b$ this equation of state is in good 
agreement with the more accurate equation of state $\beta p_{MCSL}$ 
established by Mansoori et al. \protect\cite{Mansoori71}. However, for larger 
values of $\eta$, there are deviations.}
\label{tab:sum35}
\vspace*{0.5cm}
\begin{tabular}{cccccc}
$\eta$ & $\eta_s$ & $\eta_b$ & $\sum_i \rho_i(R_i+0)$ & $\beta p_{PY}^c$ &
$\beta p_{MCSL}$ \\
\hline
0.1246 & 0.0246 & 0.1000 & 0.01804 & 0.01804 & 0.01801 \\
0.1252 & 0.0880 & 0.0372 & 0.03809 & 0.03809 & 0.03805 \\
0.1309 & 0.0026 & 0.1283 & 0.01256 & 0.01256 & 0.01253 \\
0.1850 & 0.0990 & 0.0890 & 0.06116 & 0.06116 & 0.06089 \\
0.2005 & 0.0356 & 0.1649 & 0.03901 & 0.03901 & 0.03877 \\
0.2247 & 0.0126 & 0.2121 & 0.03697 & 0.03698 & 0.03670 \\
0.2278 & 0.1812 & 0.0466 & 0.12304 & 0.12306 & 0.12202 \\
0.2670 & 0.1953 & 0.0717 & 0.16351 & 0.16354 & 0.16132 \\
0.2749 & 0.0545 & 0.2204 & 0.07985 & 0.07987 & 0.07878 \\
0.3477 & 0.0257 & 0.3220 & 0.11401 & 0.11405 & 0.11097 \\ 
0.3712 & 0.0607 & 0.3105 & 0.16908 & 0.16916 & 0.16372 \\
0.3906 & 0.0058 & 0.2984 & 0.06845 & 0.06847 & 0.06713 \\
0.3968 & 0.0120 & 0.3848 & 0.15642 & 0.15652 & 0.15018 \\
\end{tabular}
\end{table}

\begin{table}
\caption{Results for the same quantities as in Table~\ref{tab:sum35} for a 
binary hard-sphere mixture with size ratio $R_b:R_s=3:1$.}
\label{tab:sum31}
\vspace*{0.5cm}
\begin{tabular}{cccccc}
$\eta$ & $\eta_s$ & $\eta_b$ & $\sum_i \rho_i(R_i+0)$ & $\beta p_{PY}^c$ &
$\beta p_{MCSL}$ \\
\hline
0.1209 & 0.0188 & 0.1021 & 0.00778 & 0.00778 & 0.00776 \\
0.1288 & 0.0037 & 0.1251 & 0.00313 & 0.00313 & 0.00313 \\
0.1456 & 0.0283 & 0.1173 & 0.01219 & 0.01219 & 0.01219 \\
0.2230 & 0.0026 & 0.2204 & 0.00643 & 0.00643 & 0.00637 \\
0.2471 & 0.0199 & 0.2272 & 0.01590 & 0.01590 & 0.01578 \\
0.2513 & 0.0089 & 0.2424 & 0.01104 & 0.01104 & 0.01094 \\
0.3021 & 0.0016 & 0.3005 & 0.01180 & 0.01180 & 0.01159 \\
0.3257 & 0.0136 & 0.3121 & 0.02166 & 0.02166 & 0.02124 \\
0.3775 & 0.0099 & 0.3676 & 0.02837 & 0.02837 & 0.02747 \\
0.3906 & 0.0047 & 0.3859 & 0.02718 & 0.02719 & 0.02620 \\
\end{tabular}
\end{table}

\begin{table}
\caption{Individual contact values of the density profiles for a binary 
mixture with size ratio $R_b:R_s=5:3$ as obtained by the original Rosenfeld 
functional ($\sigma_i^3 \rho_i^{org}(z=R_i+0)$), the antisymmetrized 
modification with $q=3$ ($\sigma_i^3 \rho_i^{q=3}(z=R_i+0)$), and the 
interpolating modification ($\sigma_i^3 \rho_i^{int}(z=R_i+0)$) with $i=s,b$.
$\sigma_i=2 R_i$ is the diameter of species $i$. While all versions of the 
functional respect the sum rule in Eq.~(\ref{sumrule}), the individual contact
values differ slightly for the different versions of the DFT. 
$\eta=\eta_s+\eta_b$ is the total packing fraction.}
\label{tab:contact35}
\vspace*{0.5cm}
\begin{tabular}{ccccccc}
$\eta$ & $\sigma_s^3 \rho_s^{org}(R_s+)$ & $\sigma_s^3 \rho_s^{q=3}(R_s+)$ & 
$\sigma_s^3 \rho_s^{int}(R_s+)$ & $\sigma_b^3 \rho_b^{org}(R_b+)$ & 
$\sigma_b^3 \rho_b^{q=3}(R_b+)$ & $\sigma_b^3 \rho_b^{int}(R_b+)$\\
\hline
0.1246 & 0.07051 & 0.07058 & 0.07059 & 0.34182 & 0.34150 & 0.34144 \\ 
0.1252 & 0.27332 & 0.27340 & 0.27343 & 0.14522 & 0.14487 & 0.14478 \\
0.1309 & 0.00745 & 0.00746 & 0.00746 & 0.43082 & 0.43077 & 0.43077 \\
0.1850 & 0.38587 & 0.38639 & 0.38657 & 0.47860 & 0.47619 & 0.47550 \\  
0.2005 & 0.13385 & 0.13432 & 0.13444 & 0.82502 & 0.82284 & 0.82229 \\
0.2247 & 0.05002 & 0.05036 & 0.05043 & 1.13782 & 1.13628 & 1.13594 \\
0.2278 & 0.90612 & 0.90702 & 0.90727 & 0.36210 & 0.35787 & 0.35677 \\
0.2670 & 1.15640 & 1.15871 & 1.15943 & 0.70205 & 0.69099 & 0.68829 \\
0.2749 & 0.27715 & 0.27944 & 0.28026 & 1.67448 & 1.66398 & 1.65994 \\
0.3477 & 0.17094 & 0.17520 & 0.17709 & 3.43107 & 3.41124 & 3.40174 \\
0.3712 & 0.47247 & 0.48279 & 0.48746 & 4.07483 & 4.02673 & 4.00460 \\
0.3906 & 0.03099 & 0.03160 & 0.03181 & 2.39204 & 2.38927 & 2.38826 \\ 
0.3968 & 0.10016 & 0.10471 & 0.10703 & 5.33002 & 5.30870 & 5.29677 \\
\end{tabular}
\end{table}

\begin{table}
\caption{Results for the same quantities as in Table~\ref{tab:sum35} for a 
binary mixture with size ratio $R_b:R_s=3:1$.}
\label{tab:contact31}
\vspace*{0.5cm}
\begin{tabular}{ccccccc}
$\eta$ & $\sigma_s^3 \rho_s^{org}(R_s+)$ & $\sigma_s^3 \rho_s^{q=3}(R_s+)$ & 
$\sigma_s^3 \rho_s^{int}(R_s+)$ & $\sigma_b^3 \rho_b^{org}(R_b+)$ & 
$\sigma_b^3 \rho_b^{q=3}(R_b+)$ & $\sigma_b^3 \rho_b^{int}(R_b+)$\\
\hline
0.1209 & 0.04862 & 0.04864 & 0.04864 & 0.36808 & 0.36769 & 0.36761 \\
0.1288 & 0.00932 & 0.00933 & 0.00932 & 0.42401 & 0.42387 & 0.42389 \\
0.1456 & 0.07891 & 0.07895 & 0.07895 & 0.50360 & 0.50248 & 0.50231 \\ 
0.2230 & 0.00839 & 0.00842 & 0.00841 & 1.16319 & 1.16241 & 1.16248 \\
0.2471 & 0.07074 & 0.07101 & 0.07104 & 1.52497 & 1.51744 & 1.51686 \\
0.2513 & 0.03121 & 0.03136 & 0.03136 & 1.54220 & 1.53814 & 1.53814 \\
0.3021 & 0.00626 & 0.00633 & 0.00633 & 2.37903 & 2.37714 & 2.37714 \\
0.3257 & 0.05996 & 0.06072 & 0.06078 & 3.05954 & 3.03862 & 3.03744 \\
0.3775 & 0.05168 & 0.05289 & 0.05299 & 4.73342 & 4.69975 & 4.69719 \\
0.3906 & 0.02540 & 0.02612 & 0.02617 & 5.18598 & 5.16579 & 5.16450 \\
\end{tabular}
\end{table}


\begin{references}
\bibitem{eta} With $\eta_i=\frac{4 \pi}{3} R_i^3 \rho_i$ ($i=s,b$) we denote 
the packing fraction of the species $i$ in the reservoir.
\bibitem{details35} For the density profiles in Fig.~\ref{fig:prof1} we find 
the following values: $z_s^{(1)}=1.3276 \sigma_s$, $z_s^{(2)}=2.9603 \sigma_s$ 
and $\bar \rho_s^{(0)}=0.9313$, $\bar \rho_s^{(1)}=0.9719$ for the small 
spheres and $z_b^{(1)}=1.6289 \sigma_s$, $z_b^{(2)}=3.2652 \sigma_s$ and
$\bar \rho_b^{(0)}=0.8873$ and $\bar \rho_b^{(1)}=0.9589$ for the big 
spheres.
\bibitem{details31} For the density profiles in Fig.~\ref{fig:prof2} we find 
the following values: $z_s^{(1)}=2.2338 \sigma_s$, $z_s^{(2)}=5.1791 \sigma_s$ 
and $\bar \rho_s^{(0)}=1.3217$, $\bar \rho_s^{(1)}=0.9894$ for the small 
spheres and $z_b^{(1)}=3.1247 \sigma_s$, $z_b^{(2)}=6.0624 \sigma_s$ and
$\bar \rho_b^{(0)}=1.7096$ and $\bar \rho_b^{(1)}=0.9757$ for the big spheres.
\bibitem{boundaries} The precise phase boundaries towards freezing of a binary
hard-sphere mixture with size ratio $R_b:R_s=3:1$ are not known. Therefore we 
roughly estimated the onset of freezing from the phase diagram of a binary 
hard-sphere mixture with size ratio $R_b:R_s=5:1$ \cite{Dijkstra98} to be 
close to $\eta_b=0.45$ for $\eta_s \leq 0.1$, $\eta_b=0.4$ for $\eta_s=0.15$, 
$\eta_b=0.35$ for $\eta_s=0.2$, and $\eta_b=0.2$ for $\eta_s=0.25$.
\end{references}
\end{document}